\documentclass[12pt]{article}
\usepackage{eqsection,subeqnarray,indent,amsfonts,amssymb}

\begin{document}

\title{\vspace*{-3cm}
       A Personal List of Unsolved Problems \\
       Concerning Lattice Gases\\
       and Antiferromagnetic Potts Models\thanks{
   Talk presented at the conference on Inhomogeneous Random Systems,
   Universit\'e de Cergy-Pontoise, 25 January 2000.
}}

\author{
  \\[-5mm]
  {\normalsize Alan D. Sokal}                   \\[-1.5mm]
  {\normalsize \it Department of Physics}       \\[-1.5mm]
  {\normalsize \it New York University}         \\[-1.5mm]
  {\normalsize \it 4 Washington Place}          \\[-1.5mm]
  {\normalsize \it New York, NY 10003 USA}      \\[-1.5mm]
  {\normalsize \tt SOKAL@NYU.EDU}               \\[2mm]
}

\date{April 13, 2000 \\
      Version 2: May 9, 2000 \\
      Version 3: September 22, 2000 \\[1mm]
      to appear {\em in Markov Processes and Related Fields}
     }

\maketitle
\thispagestyle{empty}   

\vspace{-10mm}

\begin{abstract}
I review recent results and unsolved problems
concerning the hard-core lattice gas and the $q$-coloring model
(antiferromagnetic Potts model at zero temperature).
For each model, I consider its equilibrium properties
(uniqueness/nonuniqueness of the infinite-volume Gibbs measure,
complex zeros of the partition function)
and the dynamics of local and nonlocal Monte Carlo algorithms
(ergodicity, rapid mixing, mixing at complex fugacity).
These problems touch on mathematical physics, probability,
combinatorics and theoretical computer science.

[Pour M.~Toubon]
Je passe en revue des r\'esultats r\'ecents et des probl\`emes non r\'esolus
concernant le gaz sur r\'eseau avec exclusion
et le mod\`ele de $q$-coloriage
(mod\`ele de Potts antiferromagn\'etique \`a temp\'erature nulle).
Pour chacun des deux mod\`eles, je consid\`ere ses propri\'et\'es d'\'equilibre
(unicit\'e/non-unicit\'e de la mesure de Gibbs en volume infini,
 z\'eros complexes de la fonction de partition)
et la dynamique des algorithmes Monte Carlo locaux et non-locaux
(ergodicit\'e, m\'elange rapide, m\'elange \`a fugacit\'e complexe).
Ces probl\`emes touchent \`a la physique math\'ematique, \`a la probabilit\'e,
\`a la combinatoire et \`a l'infor\-ma\-tique th\'eorique.
\end{abstract}

\vspace{1mm}
\noindent
{\bf Running title:}  Lattice Gases and Potts Models

\vspace{3mm}

\vspace{3mm}
\noindent
{\bf AMS classification numbers:}
05C15, 05C30, 05C69, 05C85, 26C10, 60K35, 68R10, 68W20,
82B20, 82B26, 82B80, 82C20.

\clearpage

\newcommand{\be}{\begin{equation}}
\newcommand{\ee}{\end{equation}}
\newcommand{\<}{\langle}
\renewcommand{\>}{\rangle}
\newcommand{\widebar}{\overline}
\def\reff#1{(\protect\ref{#1})}
\def\spose#1{\hbox to 0pt{#1\hss}}
\def\ltapprox{\mathrel{\spose{\lower 3pt\hbox{$\mathchar"218$}}
 \raise 2.0pt\hbox{$\mathchar"13C$}}}
\def\gtapprox{\mathrel{\spose{\lower 3pt\hbox{$\mathchar"218$}}
 \raise 2.0pt\hbox{$\mathchar"13E$}}}
\def\textprime{${}^\prime$}
\def\proof{\par\medskip\noindent{\sc Proof.\ }}
\def\qed{\hbox{\hskip 6pt\vrule width6pt height7pt depth1pt \hskip1pt}\bigskip}
\def\proofof#1{\bigskip\noindent{\sc Proof of #1.\ }}
\def\half{ {1 \over 2} }
\def\third{ {1 \over 3} }
\def\twothird{ {2 \over 3} }
\def\smfrac#1#2{\textstyle{#1\over #2}}
\def\smhalf{ \smfrac{1}{2} }
\newcommand{\real}{\mathop{\rm Re}\nolimits}
\renewcommand{\Re}{\mathop{\rm Re}\nolimits}
\newcommand{\imag}{\mathop{\rm Im}\nolimits}
\renewcommand{\Im}{\mathop{\rm Im}\nolimits}
\newcommand{\Log}{\mathop{\rm Log}\nolimits}
\newcommand{\sgn}{\mathop{\rm sgn}\nolimits}
\newcommand{\spr}{\mathop{\rm spr}\nolimits}
\def\hboxscript#1{ {\hbox{\scriptsize\it #1}} }

\newcommand{\restrict}{\upharpoonright}
\newcommand{\triplenorm}{| \hspace{-0.3mm} | \hspace{-0.3mm} |}
 
\def\scra{\mathcal{A}}
\def\scrb{\mathcal{B}}
\def\scrc{\mathcal{C}}
\def\scrd{\mathcal{D}}
\def\scre{\mathcal{E}}
\def\scrf{\mathcal{F}}
\def\scrg{\mathcal{G}}
\def\scrh{\mathcal{H}}
\def\scrl{\mathcal{L}}
\def\scro{\mathcal{O}}
\def\scrp{\mathcal{P}}
\def\scrr{\mathcal{R}}
\def\scrs{\mathcal{S}}
\def\scrt{\mathcal{T}}
\def\scru{\mathcal{U}}
\def\scrv{\mathcal{V}}
\def\scrz{\mathcal{Z}}

\def\Z{{\mathbb Z}}
\def\R{{\mathbb R}}
\def\C{{\mathbb C}}

\def\bone{{\mathbf 1}}

\newtheorem{theorem}{Theorem}[section]
\newtheorem{proposition}[theorem]{Proposition}
\newtheorem{lemma}[theorem]{Lemma}
\newtheorem{corollary}[theorem]{Corollary}
\newtheorem{definition}[theorem]{Definition}
\newtheorem{conjecture}[theorem]{Conjecture}
\newtheorem{question}[theorem]{Question}
\newtheorem{problem}[theorem]{Problem}


\newenvironment{sarray}{
          \textfont0=\scriptfont0
          \scriptfont0=\scriptscriptfont0
          \textfont1=\scriptfont1
          \scriptfont1=\scriptscriptfont1
          \textfont2=\scriptfont2
          \scriptfont2=\scriptscriptfont2
          \textfont3=\scriptfont3
          \scriptfont3=\scriptscriptfont3
        \renewcommand{\arraystretch}{0.7}
        \begin{array}{l}}{\end{array}}
 
\newenvironment{scarray}{
          \textfont0=\scriptfont0
          \scriptfont0=\scriptscriptfont0
          \textfont1=\scriptfont1
          \scriptfont1=\scriptscriptfont1
          \textfont2=\scriptfont2
          \scriptfont2=\scriptscriptfont2
          \textfont3=\scriptfont3
          \scriptfont3=\scriptscriptfont3
        \renewcommand{\arraystretch}{0.7}
        \begin{array}{c}}{\end{array}}

\section{Introduction}   \label{sec1}

In this paper I propose to review some recent results
and to list some open problems
concerning a pair of statistical-mechanical lattice models
(defined precisely in Sections~\ref{sec2} and \ref{sec3}):
\begin{itemize}
   \item Hard-core lattice gas with nearest-neighbor exclusion
     and fugacity $w$ \hfill\break
     \hspace*{5mm} [= ``independent-set model'' in graph-theory language]
   \item $q$-coloring of the vertices of a graph
\end{itemize}
(These are zero-temperature models.  More generally, one can study
the hard-core lattice gas with a soft nearest-neighbor repulsion,
and the antiferromagnetic Potts model at nonzero temperature.)

What these two models have in common is that they are ``antiferromagnetic'':
nearest neighbors prefer to be (in the $q$-coloring case have to be)
in {\em different}\/ states.
Unlike ferromagnetic models, for which (roughly speaking)
only one type of ordered phase is possible,
in antiferromagnetic models the possible types
of order can depend delicately on the parameters (e.g.\ $w$ or $q$)
and on the nature of the lattice (e.g.\ bipartite or not).
This motivates studying these models on an arbitrary graph $G=(V,E)$
--- and not just on a regular lattice --- in order to investigate
which features of $G$ correlate with which features of the phase transition.
Here $G$ will be either finite or countably infinite,
depending on the question being investigated.

There is one other motivation for studying these models on arbitrary graphs:
The hard-core lattice gas on a general graph $G$
(with different fugacities $w_i \in \C$ assigned to different vertices)
is in fact the {\em universal}\/ statistical-mechanical model
in the sense that {\em any}\/ statistical-mechanical model
living on a vertex set $V_0$ can be mapped onto a gas
of nonoverlapping ``polymers'' on $V_0$,
i.e.\ a hard-core lattice gas on the intersection graph of $V_0$
\cite[Section~5.7]{Simon_93}.

For each of these models, I will discuss the following questions:
\begin{itemize}
   \item Equilibrium statistical mechanics
   \vspace{-2mm}
   \begin{itemize}
      \item Uniqueness/nonuniqueness/properties of Gibbs measures
               in infinite volume
      \item Complex zeros (in $w$ or $q$) of the partition function
               in finite volume
   \end{itemize}
   \item Monte Carlo algorithms (single-site, local and nonlocal)
   \vspace{-2mm}
   \begin{itemize}
      \item Ergodicity (this is sometimes nontrivial at zero temperature)
      \item Rapid mixing
      \item Mixing at complex $w$ or $q$ (a new and perhaps crazy idea)
   \end{itemize}
\end{itemize}
I want to stress that I'm not an expert on any of these things,
so some of my questions may be naive and some of the ``open problems''
that I pose may already be solved.

\section{Lattice Gas: Equilibrium Properties}  \label{sec2}

Let $G=(V,E)$ be a finite graph,
and let $w = \{w_x\}_{x \in V}$ be a set of (possibly complex) fugacities.
The partition function of the hard-core lattice gas
(= independent-set polynomial) on $G$ is
\be
   Z(w)   \;=\;   \sum_{\begin{scarray}
                          A \subseteq V \\
                          A \hboxscript{ independent}
                        \end{scarray}
                       }
                  \, \prod_{x \in A} w_x   \;.
 \label{eq2.1}
\ee
(Recall that $A \subseteq V$ is called an {\em independent set}\/
if it does not contain any pair of adjacent vertices.)
When the fugacities are nonnegative,
an equilibrium probability distribution on the state space $2^V$
can be defined by
\be
   \mu(A)   \;=\; Z(w)^{-1} \prod_{x \in A} w_x
                  \,\times\, \chi(A \hbox{ independent})   \;.
\ee
When $G$ is countably infinite and the fugacities are nonnegative,
Gibbs measures can then be defined by the usual
Dobrushin--Lanford--Ruelle prescription \cite{Georgii_88}.

At small $w$ we expect to have uniqueness of the Gibbs measure,
exponential decay of correlations, analyticity, etc.
Indeed, the Dobrushin uniqueness theorem \cite{Georgii_88,Simon_93}
easily implies the uniqueness of the infinite-volume Gibbs measure,
and the exponential decay of correlations in this unique Gibbs measure,
whenever $0 \le w_x \le (1-\epsilon)/(d_x - 1 + \epsilon)$ for all vertices $x$
[here $\epsilon > 0$,
 and $d_x$ denotes the number of vertices adjacent to $x$].
In particular, when all vertices are assigned the same fugacity $w$,
this happens whenever $0 \le w < 1/(\Delta-1)$
[here $\Delta = \max_{x \in V} d_x$ is the maximum degree of $G$].

A slightly better bound \cite{vandenBerg_94} can be obtained using the
disagreement-percolation method:
if the site percolation model on $G$ with occupation probabilities
$p_x = w_x/(1+w_x)$ has (with probability 1) no infinite occupied cluster,
then the hard-core lattice gas with fugacities $w$ has a unique Gibbs measure;
moreover, the two-point covariances in the lattice gas are bounded above
by the corresponding connection probabilities in the percolation model.
In particular, if $0 \le p_x \le (1-\epsilon)/(d_x - 1)$
for all vertices $x$, then standard arguments (comparison to a branching
process) show that the percolation model has (w.p.\ 1) no infinite
occupied cluster and has exponential decay of connectivities.
So we prove uniqueness and exponential decay for the lattice gas whenever
$0 \le w_x \le (1-\epsilon)/(d_x - 2 + \epsilon)$
for all vertices $x$,
or $0 \le w < 1/(\Delta-2)$ in case the fugacities are all equal.

It is natural to ask whether this result is sharp.
The answer is almost certainly no:
Vigoda \cite{Vigoda_99a,Vigoda_99b} improves $1/(\Delta-2)$ to $2/(\Delta-2)$,
at least for regular lattices,
as a byproduct of proving rapid mixing of Glauber dynamics
(see Section~\ref{sec4} below).\footnote{
   For some lattices, the disagreement-percolation bound is better than
   $2/(\Delta-2)$.  For example, the triangular-lattice site-percolation model
   has $p_c = 1/2$ \cite{Kesten_82}, which implies uniqueness of the
   triangular-lattice-gas Gibbs measure for for $w < 1$;
   but $2/(\Delta-2) = 1/2$.
}
But I suspect that this result is not sharp either,
and that the optimal result is:

\begin{conjecture}
   \label{conj2.1}
For any countably infinite graph $G$ of maximum degree $\Delta$,
the hard-core lattice gas on $G$ has a unique Gibbs measure whenever
$0 \le w_x \le (1-\epsilon) \times$\break
$(\Delta-1)^{\Delta-1}/(\Delta-2)^\Delta$
$\,[\sim e/\Delta$ for large $\Delta]$
for all vertices $x$, for some $\epsilon > 0$.
[Perhaps this is true even with $\epsilon = 0$.]
\end{conjecture}

\noindent
Conjecture~\ref{conj2.1} is motivated by the fact that
$w = (\Delta-1)^{\Delta-1}/(\Delta-2)^\Delta$
is the critical point for the complete rooted tree
with branching factor $r = \Delta -1$
\cite{Runnels_67,Kelly_85,Brightwell_99a,Brightwell_99b,Sokal_in_prep}.
It seems to be an open problem to prove Conjecture~\ref{conj2.1}
even when $G$ is a non-regular tree and the $w_x$ are all equal.

\bigskip

On a finite graph $G$, the free energy $\log Z(w)$
and the correlation functions (which are derivatives of $\log Z$)
are analytic functions of $w$ on any simply connected domain $D \subset \C$
where $Z(w)$ is nonvanishing.
Moreover, by a standard Vitali argument \cite[pp.~417--418]{Simon_93},
this analyticity can be carried over to the infinite-volume limit
whenever this limit exists for {\em real}\/ $w$
(e.g.\ for amenable transitive graphs).
So it is of great interest to find sufficient conditions
for the nonvanishing of $Z(w)$
[which is trivial when $w \ge 0$, but not when $w$ is negative or complex].
Shortly before his death, Dobrushin \cite{Dobrushin_96a,Dobrushin_96b}
proved a beautiful theorem on the absence of zeros of $Z(w)$
at small complex fugacity:

\begin{theorem}[Dobrushin \protect\cite{Dobrushin_96a,Dobrushin_96b}]
  \label{thm2.2}
Let $\{c_x\}_{x \in V}$ be an arbitrary set of nonnegative numbers,
and define
\be
   R_x   \;=\;
   (1 - e^{-c_x}) \, \exp\!\left( - \sum\limits_{y \sim x} c_y \right)
\ee
where $y \sim x$ denotes that $y$ is adjacent to $x$.
Then $Z(w)$ is nonvanishing in the closed polydisc $|w_x| \le R_x$.
\end{theorem}

\noindent
The proof is an extraordinarily simple 3-line induction on the number
of vertices in $G$:
see \cite{Dobrushin_96a,Dobrushin_96b},
or see \cite[Section 3]{Sokal_00a} for an equally simple proof
of a slightly stronger result.

\begin{corollary}
   \label{cor2.3}
If $G$ has maximum degree $\Delta$,
then $Z(w)$ is nonvanishing in the closed polydisc
$|w_x| \le \Delta^\Delta/(\Delta+1)^{\Delta+1}$
$\,[\sim 1/(e\Delta)$ for large $\Delta]$.
\end{corollary}

\noindent
{\bf Remarks.}
1. Koteck\'y and Preiss \cite{Kotecky_86} proved a result
slightly weaker than Theorem~\ref{thm2.2},
in which $1 - e^{-c_x}$ is replaced by $c_x e^{-c_x}$.

2. Corollary~\ref{cor2.3} is close to best possible:
for the complete rooted tree with branching factor $r = \Delta -1$
and depth $n$, $Z(w)$ has negative real zeros that tend to
$w = - (\Delta-1)^{\Delta-1}/\Delta^\Delta$ as $n \to\infty$
\cite{Shearer_85,Sokal_in_prep}.
Indeed, by an inductive argument closely related to
(but more subtle than) Dobrushin's,
Shearer \cite{Shearer_85} improves Corollary~\ref{cor2.3}
to the optimal radius $(\Delta-1)^{\Delta-1}/\Delta^\Delta$.
Alex Scott and I are investigating analogous improvements
of Theorem~\ref{thm2.2}.
Note that this optimal radius is a factor $\approx e$ smaller than
the interval at {\em real positive}\/ $w$ that is obtained by the
Dobrushin uniqueness theorem,
and a factor $\approx e^2$ smaller than the interval proposed in
Conjecture~\ref{conj2.1}.

3. For any lattice gas with repulsive interactions
--- and in particular for one with hard-core exclusions ---
the Mayer expansion $\log Z(w) = \sum c_{\bf n} w^{\bf n}$
has alternating signs:  $(-1)^{|{\bf n}| -1} c_{\bf n} \ge 0$.
It follows that the zero of $Z(w)$ closest to the origin
always lies on the negative real axis.

4. Theorem~\ref{thm2.2}
(which can be generalized to soft-core repulsive interactions \cite{Sokal_00a})
provides an extraordinarily simple proof
of the convergence of the Mayer expansion for such lattice gases.
Recall that the usual approach to proving convergence of the Mayer expansion
\cite{Penrose_67,Seiler_82,Cammarota_82,Brydges_86,%
Simon_93,Brydges_99,Sokal_Mayer_in_prep}
is to explicitly bound the expansion coefficients;
this requires some rather nontrivial combinatorics
(for example, an inequality of Penrose \cite{Penrose_67}
 together with the counting of trees).
Once this is done, an immediate consequence is that $Z$ is nonvanishing
in any polydisc where the series for $\log Z$ is convergent.
Dobrushin's brilliant idea
was to prove these two results in the opposite order:
first one proves, by an elementary induction on the cardinality
of the state space, that $Z$ is nonvanishing in some specified polydisc;
it then follows immediately that $\log Z$ is analytic in that polydisc,
and hence that its Taylor series is convergent there.
It is an interesting open question to know whether this approach
can be made to work without the assumption of hard-core self-repulsion.

\bigskip

What if we ask about regions of $w$-space other than
polydiscs centered at $w=0$?

\begin{question}
  \label{question2.4}
What is the largest complex domain containing $w=0$
on which $Z(w)$ is zero-free for all graphs of maximum degree $\Delta$?
[This question has two versions, depending on whether the $w_x$
are assumed equal or unequal.
In the latter version, one seeks domains $D$ for which $Z(w) \neq 0$
whenever $w_x \in D$ for all $x$;
but there need not exist a unique maximal such domain.]
\end{question}

\noindent
I conjecture that there is a complex domain $D_\Delta$ containing at least
the interval $0 \le w < 1/(\Delta-1)$ of the real axis ---
and possibly even the interval
$0 \le w < (\Delta-1)^{\Delta-1}/(\Delta-2)^\Delta$ ---
on which $Z(w)$ is zero-free for all graphs of maximum degree $\Delta$.
See also Section~\ref{sec4} regarding mixing at complex fugacity.


\bigskip

Another question is:  Can we get sharper constraints on the location of
the complex zeros of $Z(w)$ for restricted classes of graphs $G$?
A classic result of this kind is due to Heilmann and Lieb \cite{Heilmann_72}
(see also \cite{Godsil_81,Lovasz_86}):
if $G$ is a {\em line graph}\/,
then all the zeros of $Z(w)$ are {\em negative real}\/.
[The lattice gas on a line graph $L(H)$ is identical to the matching polynomial
 (monomer-dimer-model partition function) on $H$.]
Now, it is well known that $G$ is a line graph
if and only if nine particular graphs $G_1,\ldots,G_9$
do not appear as induced subgraphs of $G$ \cite[pp.~73--77]{Harary_69}.
It turns out that for eight of these ``forbidden induced subgraphs'',
all the zeros of $Z(w)$ are negative real;
the only one with non-real zeros is the {\em claw}\/ $K_{1,3} = \!\!$
\setlength{\unitlength}{5mm}
\begin{picture}(1,1)(-0.2,-0.2)
   \put(0,0){\line(1,0){2}}
   \put(0,0){\line(3,1){2}}
   \put(0,0){\line(3,-1){2}}
   \put(0,0){\circle*{0.2}}
   \put(2,0){\circle*{0.2}}
   \put(2,0.667){\circle*{0.2}}
   \put(2,-0.667){\circle*{0.2}}
\end{picture}
\hspace{1cm}
This suggests the conjecture:

\medskip

\begin{conjecture}[Hamidoune \protect\cite{Hamidoune_90},
                   Stanley \protect\cite{Stanley_98}]
   \label{conj2.5}
If $G$ is claw-free (i.e.\ has no induced subgraph $K_{1,3}$),
then all the zeros of $Z(w)$ are negative real.
\end{conjecture}

\noindent
As vague evidence in favor of this conjecture,
let us mention the following:
(a) It implies the log concavity of the counts of independent sets,
which is indeed true for claw-free graphs \cite{Hamidoune_90}.
(b) It is implied by a plausible conjecture
concerning a special symmetric function $X_G$ associated to $G$
\cite[Corollary 2.10]{Stanley_98}.
(c) The recent disagreement-percolation proof
of complete analyticity for the monomer-dimer model \cite{vandenBerg_99}
generalizes immediately to the hard-core lattice gas
on any claw-free graph \cite{vandenBerg_private}.

For possibly useful background on claw-free graphs, see \cite{Faudree_97};
see also \cite{Gasharov_99}.
For interesting results along vaguely related lines,
see \cite{Ruelle_99a,Ruelle_99b}.

\bigskip

Finally, one may ask:  What happens when the fugacity $w$ is {\em not}\/ small?
In particular, one would like to understand the uniqueness or nonuniqueness
of the infinite-volume Gibbs measure as a function of $w$,
the nature of the large-$w$ Gibbs measures,
and whether the phase transition(s) is/are first-order or second-order.
On a bipartite graph with symmetry between the two sublattices,
the mechanism driving the phase transition is clear:
at large $w$, one sublattice becomes preferentially occupied
and the other preferentially vacant.
But this does not necessarily exhaust the possible Gibbs measures:
for example, on the 3-regular tree there also exist Gibbs measures
in which the density is not constant on each sublattice
\cite{Brightwell_private}.
Moreover, the nonuniqueness of the Gibbs measure is not necessarily
monotone in $w$ \cite{Brightwell_99b}.
For general (non-bipartite) graphs, the situation is even worse,
and the nature of the possible ordered phases
seems poorly understood at present.

\section{$q$-Coloring Model: Equilibrium Properties}  \label{sec3}

Let $G=(V,E)$ be a finite graph and let $q$ be a positive integer.
A map $\sigma \colon\, V \to \{ 1,2,\ldots,q \}$
is called a {\em proper $q$-coloring}\/ of $G$
in case $\sigma(x) \neq \sigma(y)$ for all pairs of adjacent vertices
$x \sim y$.
We define $\mu$ to be the uniform probability distribution on the set
of proper $q$-colorings of $G$ (whenever this set is nonempty).
When $G$ is countably infinite,
Gibbs measures can then be defined by the usual
Dobrushin--Lanford--Ruelle prescription \cite{Georgii_88}.
The $q$-coloring problem is the zero-temperature limit of the
antiferromagnetic $q$-state Potts model \cite{Wu_82,Wu_84}.

When $q$ is large compared to the maximum degree $\Delta$,
the conditional probability distribution of $\sigma(x)$
depends only weakly on $\{\sigma(y)\}_{y \sim x}$,
so we expect to have uniqueness of the infinite-volume Gibbs measure
and exponential decay of correlations.
This is in fact the case:

\begin{theorem}[Koteck\'y \protect\cite{Kotecky_88},
                Salas and Sokal \protect\cite{Salas_97}]
  \label{thm3.1}
If $q > 2\Delta$, the hypotheses of the Dobrushin uniqueness theorem hold.
Consequently, there is a unique infinite-volume Gibbs measure,
and it has exponential decay of correlations.
\end{theorem}

\noindent
More generally, for antiferromagnetic Potts models with $q > 2\Delta$,
there is a unique infinite-volume Gibbs measure,
with exponential decay of correlations,
uniformly down to and including zero temperature \cite{Salas_97}.
Otherwise put, zero temperature belongs to the high-temperature regime!
The reason, of course, is the large ground-state entropy.

Salas and Sokal \cite{Salas_97} have some improvements of this result
for specific regular lattices (square, triangular, hexagonal, Kagom\'e),
based on decimation and a computer-assisted proof.
But their results are far from sharp.

\begin{problem}
   \label{prob3.2}
Prove uniqueness of the infinite-volume Gibbs measure
and exponential decay of correlations for
$q=3$ on the hexagonal lattice,
$q=4,5,6$ on the square lattice, etc.
\end{problem}

\noindent
For the $q=4$ square-lattice model,
perhaps one could use the Ashkin-Teller representation
plus antipercolation ideas.

It is natural to ask whether the borderline $2\Delta$ is sharp
for some (perhaps irregular) lattice.
The answer is most likely no:  Vigoda \cite{Vigoda_00} improves $2\Delta$
to ${11 \over 6} \Delta$, at least for regular lattices,
as a byproduct of proving rapid mixing of a particular local dynamics
(see Section~\ref{sec5} below).
But I suspect that this result is not sharp either.
Even for the $\Delta$-regular tree, the correct borderline is not known;
all that is known is that
there is a nonunique Gibbs measure whenever $q \le \Delta$
\cite{Peruggi_87,Brightwell_00a,Brightwell_00b}
and a unique Gibbs measure whenever $q \ge$ a certain $q_c(\Delta)$
that satisfies $q_c(\Delta) \le {5 \over 3}\Delta-1$
and $\limsup_{\Delta\to\infty} q_c(\Delta)/\Delta \le 1.6296$
\cite{Brightwell_00b}.
For trees the Gibbs measure may well be unique as soon as $q > \Delta$;
and it is at least conceivable that this is the sharp borderline
for general graphs.

\bigskip

By the Fortuin-Kasteleyn representation \cite{Kasteleyn_69,Fortuin_72},
the $q$-state Potts model can be analytically continued to {\em complex}\/ $q$;
indeed, on any finite graph the Potts-model partition function is a
{\em polynomial}\/ in $q$
(see e.g.\ \cite[Section 1]{Sokal_00a} for a review).
Specializing to the $q$-coloring problem, we conclude that
$P_G(q) = \hbox{\#(proper $q$-colorings of $G$)}$
is the restriction to $\Z_+$ of a polynomial in $q$,
which is called the {\em chromatic polynomial}\/ of $G$
\cite{Read_68,Read_88,Chia_97}.
It therefore makes sense to study the real or complex zeros of $P_G(q)$.
Here are some recent results:

\begin{theorem}[Sokal \cite{Sokal_00a}]
  \label{thm3.3}
\quad\vspace{-2mm}\par
\begin{itemize}
   \item[(a)] For a graph $G$ of maximum degree $\Delta$,
all the (real or complex) zeros of $P_G(q)$ lie in the disc $|q| < C(\Delta)$,
where $C(\Delta)$ is the solution of a certain transcendental equation
and satisfies $C(\Delta) < 7.963907 \Delta$.
   \item[(b)]  For a graph $G$ of second-largest degree $\Delta$,
all the (real or complex) zeros of $P_G(q)$ lie in the disc
$|q| < C(\Delta) +1$.
\end{itemize}
\end{theorem}

\noindent
The proof is based on writing the Fortuin-Kasteleyn representation of $P_G(q)$
as a gas of nonoverlapping polymers $P$ with fugacity $w(P) \sim q^{-(|P|-1)}$.
When $|q|$ is large enough, one can apply Theorem~\ref{thm2.2} 
to this polymer gas, and conclude that $Z \neq 0$.

\begin{question}
   \label{question3.4}
\nopagebreak
\quad\vspace{-2mm}\par\nopagebreak
\begin{itemize}
   \item[(a)]  What is the sharp bound $C_{opt}(\Delta)$?
      Might $2\Delta$ work?  Is $C_{opt}(3) = 3$?
      What about regions of the $q$-plane other than discs centered at $q=0$?
   \item[(b)]  Can ``second-largest degree'' be improved to ``max flow''?
      (See \cite[Section 7]{Sokal_00a} for details.)
   \item[(c)]  Can one obtain a {\em sublinear}\/ bound for suitable
      subclasses of graphs?
   \vspace{-2mm}
   \begin{itemize}
      \item For theta graphs $\Theta_{s_1,\ldots,s_p}$
         [consisting of a pair of endvertices connected by
          $p$ internally disjoint paths of lengths $s_1,\ldots,s_p$]
         the answer is {\em yes}\/:  there is a bound of order $p/\log p$
         \cite{Brown_et_al}.
      \item For series-parallel graphs, I conjecture that the answer is
         {\em yes}\/ (again $p/\log p$) for maximum degree,
         and {\em no}\/ for second-largest degree or max flow;
         I am currently trying to prove these claims.
      \item For planar graphs, it is an open question.
   \end{itemize}
\end{itemize}
\end{question}

Let $C$ be the (countably infinite) set consisting of the chromatic roots
of all finite graphs, and let $\overline{C}$ be its closure.
Jason Brown (unpublished) asked whether $\overline{C}$
has nonzero two-dimensional Lebesgue measure.
The answer is yes, in the strongest possible sense:
$\overline{C}$ is the whole complex plane!
More precisely, one has:

\begin{theorem}[Sokal \cite{Sokal_00b}]
   \label{thm3.5}
For theta graphs $\Theta^{(s,p)}$
[= $p$ parallel-connected chains each consisting of $s$ edges in series],
the roots of their chromatic polynomials, taken together,
are dense in the entire complex $q$-plane
with the possible exception of the disc $|q-1| < 1$.
\end{theorem}

\noindent
For a brief discussion of the intuition behind the proof,
see \cite{Sokal_le70}.

\begin{corollary}[Sokal \cite{Sokal_00b}]
   \label{cor3.6}
There is a countably infinite family of (not-necessarily-planar)
graphs whose chromatic zeros are, taken together,
dense in the entire complex $q$-plane.
\end{corollary}

\noindent
Indeed, it suffices to consider the union of the two families
$\Theta^{(s,p)}$ and $\Theta^{(s,p)} + K_2$
and to recall that
$P_{G + K_n}(q) = q(q-1) \cdots (q-n+1) \, P_G(q-n)$.\footnote{
   Here $K_2 = \!\!$
   \setlength{\unitlength}{5mm}
   \begin{picture}(1,0.4)(-0.2,-0.2)
      \put(0,0){\line(1,0){2}}
      \put(0,0){\circle*{0.2}}
      \put(2,0){\circle*{0.2}}
   \end{picture}
   \hspace{6mm}
   is the complete graph on two vertices.
   The {\em join}\/ of two graphs, denoted $G_1 + G_2$,
   is obtained from the disjoint union $G_1 \cup G_2$
   by adjoining an edge between each pair of vertices
   $x \in G_1$ and $y \in G_2$.
}

\section{Lattice Gas: Dynamics of Algorithms}  \label{sec4}

The single-site heat-bath (Glauber) dynamics
for the hard-core lattice gas on a finite graph $G=(V,E)$
is defined as follows:
Choose uniformly at random a vertex $x \in V$;
make $x$ occupied with probability $w_x / (1+w_x)$
if all the neighbors of $x$ are vacant,
and make $x$ vacant otherwise.
This is easily seen to define an irreducible Markov chain
satisfying detailed balance with respect to the Gibbs distribution $\mu$.
We want to know:  Under what conditions is this Markov chain rapidly mixing,
in the sense that its mixing time\footnote{
   We use here the computer scientists' definition of mixing time:
   Let
   $$
      \Delta_x(t)  \;=\; {1 \over 2} \sum\limits_y | (P^t)_{xy} - \pi_y |
   $$
   be the distance to stationarity after time $t$, starting in state $x$;
   and define the (worst-case) mixing time
   $$
      \tau(\epsilon)   \;=\;
      \max\limits_x \min \{ t \colon\; \Delta_x(t') \le \epsilon
                            \hbox{ for all } t' \ge t \}
      \;.
   $$
   Other quantities of interest are the autocorrelation times
   $\tau_{{\rm int}, A}$ and $\tau_{\rm exp}$ defined in
   \cite{Sokal_Cargese};  in particular, $\tau_{\rm exp}$ is
   essentially the inverse spectral gap.
   [More precisely, $\exp(-1/\tau_{\rm exp}) = \spr(P)$,
    where $\spr(P)$ is the spectral radius of $P$
    acting on the quotient space modulo constant functions.]
}
is bounded by a (hopefully low-order) polynomial in $n$
(the number of vertices in $G$)?\footnote{
   One can also consider more general local dynamics,
   such as the single-edge heat-bath algorithm \cite{Luby_97,Dyer_00}
   or the Dyer-Greenhill chain \cite{Dyer_00}.
}

The standard proof of the Dobrushin uniqueness theorem
(see e.g.\ \cite[Section~5.1]{Simon_93}) shows also the $O(n \log n)$
mixing of the Glauber dynamics whenever the Dobrushin hypothesis holds.
In particular, this holds for the hard-core lattice gas
whenever $0 \le w < 1/(\Delta-1)$.
This result can alternatively be derived by path-coupling arguments,
which also show $O(n^2 \log n)$ mixing at $w = 1/(\Delta-1)$ \cite{Dyer_00}.

These results are, however, not optimal.
By more intricate coupling arguments,
Luby and Vigoda \cite{Luby_99,Vigoda_99a,Vigoda_99b}
have recently proven $O(n \log n)$ mixing of the Glauber dynamics
for $0 \le w < 2/(\Delta-2)$,
and $O(n^3)$ mixing at $w = 2/(\Delta-2)$.\footnote{
   Dyer and Greenhill \cite{Dyer_00}
   earlier proved $O(n \log n)$ mixing when $0 \le w < 2/(\Delta-2)$,
   and $O(n^2 \log n)$ mixing when $w = 2/(\Delta-2)$,
   for a slightly more complicated dynamics that acts on
   a vertex $v$ {\em and its neighbors}\/.
   By a comparison argument (see also \cite{Randall_00}),
   they then deduced $O(n^3 \log n)$ mixing
   when $0 \le w < 2/(\Delta-2)$,
   and $O(n^4 \log n)$ mixing when $w = 2/(\Delta-2)$,
   for the Glauber dynamics.
   But the Luby-Vigoda direct analysis appears to give better bounds.
}

The mixing properties of the Glauber (or any local) dynamics
are closely related to the model's equilibrium properties.
Indeed, consider a model defined on the regular lattice $\Z^d$,
and suppose that there exists a local dynamics
with mixing time $O(L^{d+1-\epsilon})$ on a cube of side $L$,
uniformly in the boundary condition.
Then van den Berg \cite{vandenBerg_unpub}
shows that there is a unique infinite-volume Gibbs measure.
Conversely, suitable conditions of ``weak dependence on boundary conditions''
--- which are stronger than uniqueness of the Gibbs measure,
but which are implied by most uniqueness proofs ---
imply $O(L^d \log L)$ mixing of the Glauber dynamics,
uniformly in the boundary condition \cite{Martinelli_97}.

I'd like to raise a general question:
What is the relation between proofs of rapid mixing by coupling methods
and by the Dobrushin uniqueness method?
Otherwise put, suppose we try to translate the coupling proof
of rapid mixing from probabilistic to analytic language:
we conclude that the transition matrix $P$
(acting on the quotient space modulo constant functions)
has spectral radius $\le \beta < 1$,
hence has norm $\le \beta + \epsilon < 1$ {\em in some norm or other}\/
--- but {\em which}\/ norm?
I can give an answer in the usual situation where
the coupling is Markovian, with a transition matrix $Q(x,y \to x',y')$
satisfying $\sum_{y'} Q(x,y \to x',y') = P(x \to x')$ for all $y$
and $\sum_{x'} Q(x,y \to x',y') = P(y \to y')$ for all $x$,
and where there is a Lyapunov function $\Phi(x,y) \ge 0$ satisfying
\be
   E\Bigl( \Phi(X_{t+1},Y_{t+1}) \,\Bigl|\, X_t,Y_t \Bigr)
   \;\le\;
   \beta \Phi(X_t,Y_t)
   \;.
\ee
Then a simple calculation shows that
\be
   \| Pf \| _{{\rm Lip}(\Phi)}   \;\le\;   \beta \| f \| _{{\rm Lip}(\Phi)}
   \;,
\ee
where the Lipschitz seminorm is defined by
\be
   \| f \| _{{\rm Lip}(\Phi)}   \;=\;
   \sup\limits_{x \neq y}  {|f(x) - f(y)|  \over  \Phi(x,y)}
   \;.
\ee
This approach could perhaps be useful in trying to understand
the circumstances under which the coupling method
can or cannot give sharp bounds \cite{Burdzy_00}.

My reasons for wanting to translate a beautiful probabilistic proof
into ugly analytic language are linked to my next topic,
which is the mixing properties of Glauber dynamics at
{\em complex}\/ fugacity $w$.
(I know this sounds crazy, but please bear with me.)
Since the elements of the Glauber transition matrix $P$
are rational functions of $w$,
we can certainly extend them to complex $w$.
The matrix $P$ continues to satisfy $P\bone = \bone$,
but its rows $P(x, \,\cdot\,)$ are complex measures
with total variation norm $> 1$ whenever $w$ is negative or complex.
Nevertheless, it is easy to prove the following result:

\begin{theorem}[Sokal \protect\cite{Sokal_in_prep}]
   \label{thm4.1}
Let $G$ be a finite graph, and suppose that at fugacity $w \in \C$
the Glauber transition matrix $P$ has spectral radius $< 1$
(acting on the quotient space modulo constant functions).
Then $Z(w) \neq 0$.
\end{theorem}

\noindent
This theorem thus provides a technique for proving the nonvanishing
of $Z(w)$ that is alternative to Theorem~\ref{thm2.2},
and is inspired in part by old work of Israel \cite{Israel_76}.

The problem is:  How to verify that $\spr(P) < 1$?
One might hope to do this adapting the standard proof
of the Dobrushin uniqueness theorem (see e.g.\ \cite{Simon_93});
but surprisingly, this proof {\em falls apart immediately}\/
as soon as $w$ is even slightly negative or complex,
if one employs (as usual) the total oscillation seminorm.
(The proof fails in the strangest way:  ``dusting'' a site $i$ can cause the
``dirt'' on {\em distant}\/ sites $j$ to be amplified exponentially!)
But this does not mean that the Dobrushin uniqueness {\em method}\/
is failing;  it may mean only that we are not using the right seminorm.
Indeed, for any finite graph $G$ we clearly have $\spr(P) < 1$
for all $w \in [0,\infty)$, so by continuity this must hold also
in some ($G$-dependent) complex neighborhood of $[0,\infty)$.
Which one?  In particular, can we find a complex domain
in which $\spr(P) < 1$ holds for all graphs $G$ of maximum degree $\Delta$?

\begin{conjecture}
   \label{conj4.2}
There exists a complex domain $D_\Delta$ containing the interval\break
$[0, \, 1/(\Delta-1) \, )$
such that $\spr(P) < 1$ holds for all graphs $G$ of maximum degree $\Delta$
whenever $w \in D_\Delta$.
[As with Question~\ref{question2.4}, there are two versions of this conjecture,
depending on whether or not the $w_x$ are assumed equal.]
\end{conjecture}

\noindent
Indeed, it is not out of the question that $1/(\Delta-1)$
can be replaced here by $2/(\Delta-2)$ or even by
$(\Delta-1)^{\Delta-1}/(\Delta-2)^\Delta$.
Unfortunately, what I am thus far able to prove is much weaker, namely:

\begin{theorem}[Sokal \protect\cite{Sokal_in_prep}]
   \label{thm4.3}
Let $\{c_x\}_{x \in V}$ be an arbitrary set of positive numbers,
and define
\be
   \widetilde{R}_x   \;=\;
   (e^{c_x} - 1) \, \exp\!\left( - \sum\limits_{y \sim x} c_y \right)
\ee
where $y \sim x$ denotes that $y$ is adjacent to $x$.
Then $\spr(P) < 1$ holds in the open polydisc
$|w_x/(1+w_x)| < \widetilde{R}_x$.
In particular, $Z(w)$ is nonvanishing there.
\end{theorem}

\begin{corollary}[Sokal \protect\cite{Sokal_in_prep}]
   \label{cor4.4}
If $G$ has maximum degree $\Delta$,
then $\spr(P) < 1$ and $Z(w) \neq 0$ hold in the open polydisc
\be
   \left| {w_x \over 1+w_x} \right|
   \;<\;
   {(\Delta-1)^{\Delta-1}  \over  \Delta^\Delta}
   \;.
 \label{eqcor4.4}
\ee
\end{corollary}

\noindent
Theorem~\ref{thm4.3} is strikingly similar to Theorem~\ref{thm2.2},
but is strictly stronger [since $\widetilde{R}_x/(1 + \widetilde{R}_x) > R_x$].
Corollary~\ref{cor4.4} is very close to
Shearer's improvement of Corollary~\ref{cor2.3},
but is slightly weaker (resp.\ stronger) than Shearer's result
when $w < 0$ (resp.\ $w > 0$);
it is thus almost but not quite sharp for $w < 0$.
The proof of Theorem~\ref{thm4.3} uses an
exponentially weighted seminorm on ``Fourier coefficients''
that is completely different from the total oscillation seminorm:
Let $\eta_x = 0,1$ be the occupation variable at site $x \in V$;
write $f(\eta) = \sum\limits_{X \subseteq V} a_X
                 \prod\limits_{x \in X} (1-\eta_x)$
and define
\be
   \triplenorm f \triplenorm_{\bf c}  \;=\;
   \sum_{\begin{scarray}
            X \subseteq V \\
            X \neq \emptyset
         \end{scarray} }
   \exp\!\left( \: \sum\limits_{x \in X} c_x \right) \, |a_X|
\ee
for each nonnegative vector ${\bf c} = \{c_x\}_{x \in V}$.
The question is:  How to interpolate smoothly between these two
radically different seminorms,
which work in different regions of $w$-space, in order to prove
$\spr(P) < 1$ for a domain $D_\Delta$ containing {\em both}\/
the disc \reff{eqcor4.4} and the interval $[0, \, 1/(\Delta-1) \, )$?

\section{$q$-Coloring Model: Dynamics of Algorithms}  \label{sec5}

The single-site heat-bath (Glauber) dynamics
for the $q$-coloring model on a finite graph $G=(V,E)$ is defined as follows:
Choose uniformly at random a vertex $x \in V$,
and give $\sigma(x)$ a new value (independent of the old one)
from the uniform distribution over all colors different from
$\{ \sigma(y) \}_{y \sim x}$.
More generally, one can consider the single-edge heat-bath dynamics
in which a pair of adjacent sites are simultaneously recolored
conditional on their neighbors \cite{Dyer_98},
or a yet more complicated local dynamics
(e.g.\ \cite{Bubley_98,Vigoda_00}).
Finally, one can consider the Wang-Swendsen-Koteck\'y (WSK)
nonlocal cluster dynamics \cite{WSK_89,WSK_90}, which is defined as follows:
Choose uniformly at random a pair of distinct colors
$\alpha,\beta \in \{1,\ldots,q\}$;
let $G_{\alpha\beta}$ be the induced subgraph of $G$
consisting of sites $x$ for which $\sigma(x) = \alpha$ or $\beta$;
then, independently on each connected component of $G_{\alpha\beta}$,
with probability ${1 \over 2}$ either leave that component alone
or else interchange colors $\alpha$ and $\beta$ on it.

All these Markov chains are easily seen to satisfy detailed balance
with respect to the uniform distribution over proper $q$-colorings.
However, since we are at zero temperature,
ergodicity (= irreducibility) is already a nontrivial question.
The following results are known:
\begin{itemize}
   \item Single-site dynamics is ergodic for $q \ge \Delta +2$ \cite{Jerrum_95},
      and can be nonergodic for $q = \Delta +1$ (consider $G = K_{\Delta+1}$)
   \item Single-edge dynamics is ergodic for $q \ge \Delta +1$ \cite{Dyer_98},
      and can be nonergodic for $q = \Delta$
      (consider $G = K_{\Delta+1}$ minus a single edge)
   \item WSK dynamics is ergodic for $q \ge \Delta +1$ \cite{Jerrum_private}
   \item WSK dynamics is ergodic for all $q$ if $G$ is bipartite
      \cite{Burton_Henley_97,Ferreira-Sokal}
   \item WSK dynamics is ergodic for $q=4$ on subsets of the triangular lattice
      with {\em free}\/ boundary conditions \cite{Moore_99}
   \item WSK dynamics is {\em non}\/ergodic for $q=3$ on $3m \times 3n$
      square lattices with {\em periodic}\/ boundary conditions,
      whenever $m,n$ are relatively prime \cite{Lubin_93}
   \item WSK dynamics is {\em non}\/ergodic for $q=4$ on the $6 \times 6$
      triangular lattice with {\em periodic}\/ boundary conditions,
      and for $q=3$ on the $3 \times 3$ Kagom\'e lattice with {\em periodic}\/
      boundary conditions\footnote{
   By a ``$3 \times 3$ Kagom\'e lattice'' I mean one that is obtained from
   a $3 \times 3$ triangular lattice (= the Bravais lattice of the Kagom\'e)
   by placing 3 Kagom\'e sites on each triangular site.
   It thus has $3 \times 3 \times 3 = 27$ sites.
}
      \cite{Salas_unpub}
   \item Whenever $q \le \Delta/2$ and $q$ is sufficiently large,
      there exists a graph of maximum degree $\Delta$ and
      chromatic number $O(q/\log q)$ on which the WSK dynamics
      for $q$-colorings is {\em non}\/ergodic \cite{Luczak_00}
\end{itemize}
All other cases are open (to the best of my knowledge).
In particular, we would like to know for which values of $n$ (if any)
the WSK dynamics is ergodic for $q=4$ on an $n \times n$ periodic
triangular lattice, or for $q=3$ on an $n \times n$ Kagom\'e lattice.\footnote{
   Numerical experiments \cite{Salas_unpub} suggest that WSK dynamics is
   nonergodic for $q=3$ also on the $6 \times 6$ Kagom\'e lattice
   (i.e.\ the one with $6 \times 6 \times 3 = 108$ sites).
}

When the algorithm is ergodic, the next step is to investigate
its mixing rate.  Here is what is known:

(a) For the single-site heat-bath dynamics, the Dobrushin uniqueness
argument proves $O(n \log n)$ mixing whenever $q > 2\Delta$ \cite{Salas_97}.
This result can alternatively be derived by coupling arguments,
which also show $O(n^3)$ mixing at $q = 2\Delta$ \cite{Jerrum_95,Dyer_98}.
More generally, these results hold for the antiferromagnetic Potts model
at inverse temperature $\beta$ in case $q > 2\Delta (1-e^{-\beta})$
\cite{Jerrum_95}.

(b) For the single-edge heat-bath dynamics, a path-coupling argument
shows $O(n \log n)$ mixing whenever $q \ge 2\Delta$ \cite{Dyer_98}.

(c) For a local algorithm involving simultaneous recoloring of a site
and all its neighbors, a path-coupling argument shows $O(n \log n)$ mixing
for $q=5$ when $\Delta=3$,
and for $q=7$ when $G$ is a 4-regular triangle-free graph \cite{Bubley_98}.
This shows that the $2\Delta$ bound can be beaten.
By a comparison argument, one also obtains rapid mixing of the
single-site heat-bath dynamics,
but the bounds are poor [$O(n^6 \log n)$ and $O(n^7 \log n)$, respectively].

(d) For a local algorithm involving WSK moves on components of
size $\le 6$ only, a path-coupling argument shows $O(n \log n)$ mixing for
$q > {11 \over 6} \Delta$,
and polynomial-time mixing at $q = {11 \over 6} \Delta$ \cite{Vigoda_00}.
By a comparison argument, one also obtains
$O(n^2 \log n)$ mixing for $q > {11 \over 6} \Delta$,
and polynomial-time mixing at $q = {11 \over 6} \Delta$,
for the single-site heat-bath dynamics
and for the single-cluster variant of the full WSK dynamics
\cite{Vigoda_00}.

(e) For the full WSK algorithm,
numerical data strongly suggest that there is constant-time mixing
(i.e.\ the complete absence of critical slowing-down)
for the $q=3$ model on $n \times n$ periodic square lattices
(at least for $n$ even),
uniformly down to and including the zero-temperature critical point
\cite{Salas_98,Ferreira-Sokal}.
It would be wonderful to be able to {\em prove}\/ something!

(f) \L{}uczak and Vigoda \cite{Luczak_00} construct a family $F_n$
of {\em planar}\/ (hence 4-colorable!)\ graphs
on which the WSK algorithm is torpidly mixing
[i.e.\ has mixing time $\gtapprox \exp(n^\delta)$ for some $\delta > 0$]
for arbitrarily large $q$
[indeed, for any $q = O(n^{1-\epsilon})$].
They also show that for each pair $(q,\Delta)$ satisfying
$3 \le q < \Delta/(20 \log\Delta)$,
there exists a family $G_{q,\Delta,n}$ of {\em bipartite}\/ graphs
of maximum degree $\Delta$ on which the WSK algorithm is torpidly mixing.

\section*{Acknowledgments}

I thank Jason Brown, Graham Brightwell, \"Olle H\"aggstr\"om,
Mark Jerrum, Jes\'us Salas, Alex Scott and Eric Vigoda
for many helpful conversations.
I also thank an anonymous referee for helpful comments
on an earlier draft of this paper.

This research was supported in part by
U.S.\ National Science Foundation grant PHY--9900769.
It was carried out during a Visiting Fellowship at All Souls College, Oxford,
where it was supported in part by Engineering and Physical Sciences
Research Council grant GR/M 71626
and aided by the warm hospitality of John Cardy and the
Department of Theoretical Physics.


\end{document}